\documentclass[10pt,twocolumn,pre,floatfix,superscriptaddress]{revtex4}
\usepackage{color} 
\usepackage{tabularx} 
\usepackage{amsmath} 
\usepackage{amssymb} 
\usepackage{graphicx}
\usepackage{wasysym}

\usepackage[cmintegrals,cmbraces]{newtxmath}
\usepackage{ebgaramond-maths}
\usepackage[T1]{fontenc}

\begin{document}

\title{\it Do tensor renormalization group methods work for frustrated spin
systems?}

\author{Zheng Zhu}
\affiliation {Department of Physics and Astronomy, Texas A\&M University,
College Station, Texas 77843-4242, USA}

\author{Helmut G. Katzgraber}
\affiliation{Microsoft Quantum, Microsoft, Redmond, WA 98052, USA}
\affiliation{Department of Physics and Astronomy, Texas A\&M University, College Station, Texas 77843-4242, USA}
\affiliation{Santa Fe Institute, 1399 Hyde Park Road, Santa Fe, New Mexico 87501 USA}

\date{\today}

\begin{abstract}

No. To illustrate that tensor renormalization group methods are poorly
suited for frustrated magnetic systems, we study the thermodynamic
properties of the two-dimensional Edwards-Anderson Ising spin-glass
model on a square lattice.  We show that the limited precision of
standard 64-bit data types and not a small cut-off parameter is the main
reason for unphysical negative partition function values in spin
glasses.

\end{abstract}

\maketitle

The simulation of spin-glasses \cite{binder:86} with nearest neighbor
interactions remains numerical challenge.  To date---with only few
exceptions \cite{thomas:07,wang:01}---the thermodynamic behavior of spin
glasses is best studied using Markov-chain Monte Carlo methods paired
with accelerators, such as cluster updates \cite{zhu:15b} or tempering
\cite{hukushima:96}.  Therefore, finding an algorithm that outperforms
Monte Carlo for large-scale simulations at low temperatures where the
dynamics of spin glasses is exponentially slow is of much interest.

In classical statistical system with local interactions, the Boltzmann
weight can be expressed as a tensor product. Moreover, all thermodynamic
quantities can be determined by studying an equivalent tensor-network
models. The tensor renormalization group method (TRG) is a real-space
renormalization group approach initially introduced by Levin and Nave
\cite{levin:07} for classical spin systems on two-dimensional regular
lattices. Later, the approach was further improved by Xie {\em et
al.}~\cite{xie:12} and successfully applied to the two- and
three-dimensional ferromagnetic Ising model.

Unlike the ferromagnetic Ising model, the Edwards-Anderson Ising
spin glass is a magnetic system exhibiting both quenched disorder and
frustration, i.e., it has no translation symmetry. Whether the TRG
method can be applied to the Edwards-Anderson Ising spin glass has
recently been investigated by Wang {\em et al.}~\cite{wang:13c}.  They
find that the TRG method might lead to negative values in the partition
function at low temperature and argue that a larger cut-off parameter
could be used to alleviate this problem.  In this note we argue that the
primary reason for negative partition function terms is the limited
precision of the data type (\texttt{double}) used, instead of a small cut-off
parameter. We show that TRG fails because of the near-cancellation of
the positive and negative tensor components in the partition function. A
high-precision data type is thus required to study spin glasses using 
TRG methods.\newline

\noindent {\em Model} --- The Hamiltonian of the Edwards-Anderson Ising
spin glass is given by ${\mathcal H}(\{s_i\})=-\sum_{\langle
i,j\rangle}^N J_{ij} s_i \,s_j$, where the spins $s_i \in\{\pm
1\}$ are on a square lattice and the summation is over nearest
neighbors. We use bimodal-distributed interactions between the spins,
i.e., ${\mathcal P}(J_{ij}) =  p\delta(J_{ij}-1) +
(1-p)\delta(J_{ij}+1)$.  where $p$ is the fraction of ferromagnetic
bonds.

\noindent {\em Algorithm} --- The partition function of a classical
statistical mechanical model with local interactions can be obtained by
taking the product of local tensors at each site and summing over all
bond indices, i.e., 
$Z={\rm Tr}\prod_{i}{T_{x_{i}x'_{i}y_{i}y'_{i}}}$.
The local tensor $T_{x_{i}x'_{i}y_{i}y'_{i}}$ can be defined by tracing 
out $s_{i}$ from the product matrices $W$, i.e., 
\begin{equation}
{\mathcal T_{x_{i}x'_{i}y_{i}y'_{i}}}=\sum\limits_{\alpha}{W^{2}_{\alpha,x_{i}}W^{1}_{\alpha,x'_{i}}W^{2}_{\alpha,y_{i}}W^{1}_{\alpha,y'_{i}}}
\end{equation}
 where $W^{1}$ and $W^{2}$ are $2\times2$ matrices defined by 
\begin{equation}
\!\!\!\!\!\!\!\!\!\!\!\!\!\!\!\!\!\!\!\!\!\!\!\!\!\!\!\!\!\!\!\!{\mathcal W^{1} = \quad \left( \begin{array}{ccc}
\sqrt{\cosh(1/T)} & \sqrt{\sinh(1/T)}\\
\sqrt{\cosh(1/T)} &-{ \sqrt{\sinh(1/T)}} \end{array} \right)},
\end{equation}

\vspace*{-1.0em}

\begin{equation}
{\mathcal W^{2}= \left\{ \begin{array}{llll}
\begin{pmatrix}
\sqrt{\cosh(1/T)} & \sqrt{\sinh(1/T)}\\
\sqrt{\cosh(1/T)} &-{ \sqrt{\sinh(1/T)}}
\end{pmatrix} 
& \mbox{if $J_{ij} =1$},
\\
{}
\\
\begin{pmatrix}
\sqrt{\cosh(1/T)} & -{\sqrt{\sinh(1/T)}}\\
\sqrt{\cosh(1/T)} & { \sqrt{\sinh(1/T)}}
\end{pmatrix} 
& \mbox{if $J_{ij} =-1$}.
\end{array} \right. }
\end{equation}
To coarse grain the network, two neighboring tensors are contracted
into one, i.e., 
\begin{equation}
T^{(n+1)}_{x_{1}x_{2},x'_{1}x'_{2},y,y'}=\sum\limits_{i}{T^{(n)}_{x_{1},x'_{1},y,i}T^{(n)}_{x_{2},x'_{2},i,y'}}.
\end{equation}
The lattice size is reduced by a factor of $2$ and the contracted tensor
$T^{n+1}$ (along $x$ or $y$ axis) has a higher bond dimension $D^{2}$. 
The TRG method
then truncates tensor $T^{n+1}$ into a lower rank tensor using different
strategies \cite{xie:12}.\newline

\begin{figure}[!tb]
\includegraphics[width=0.8\columnwidth]{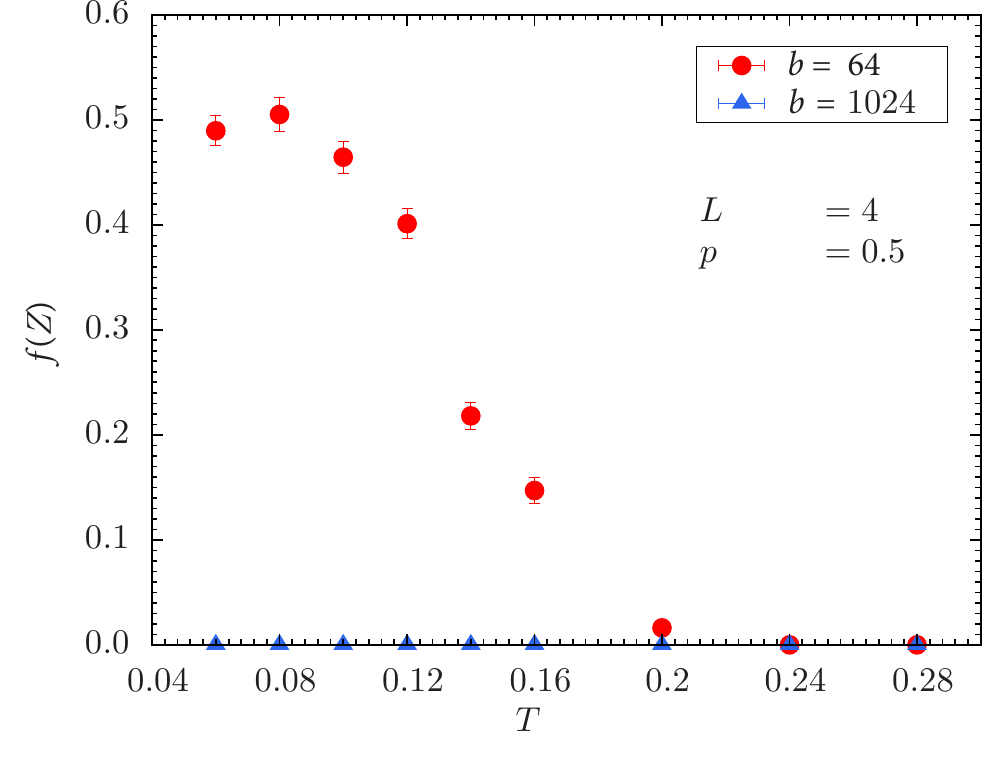}
\vspace*{-1.5em}
\caption{\small
Failure rate of the partition function $f(Z)$ of the two-dimensional
Ising spin glass as function of temperature $T$. Data
for $960$ samples for a system with $N = 16$ ($L = 4$) spins and $p =
0.50$.  The error bars represent a disorder average over the $960$
samples.
}
\label{fig1}
\end{figure}

\noindent {\em Results} --- In our C++ implementation all quantities are
stored with {\texttt double} or higher precision data types using the
GNU Multiple Precision Arithmetic Library (\texttt{gmplib.org}). Without the TRG
approximation, the coarse graining scheme based on a tensor-network
model should produce the exact partition function of the model with $Z >
0$ for all temperatures $T$.

\begin{figure}[!tb]
\includegraphics[width=0.8\columnwidth]{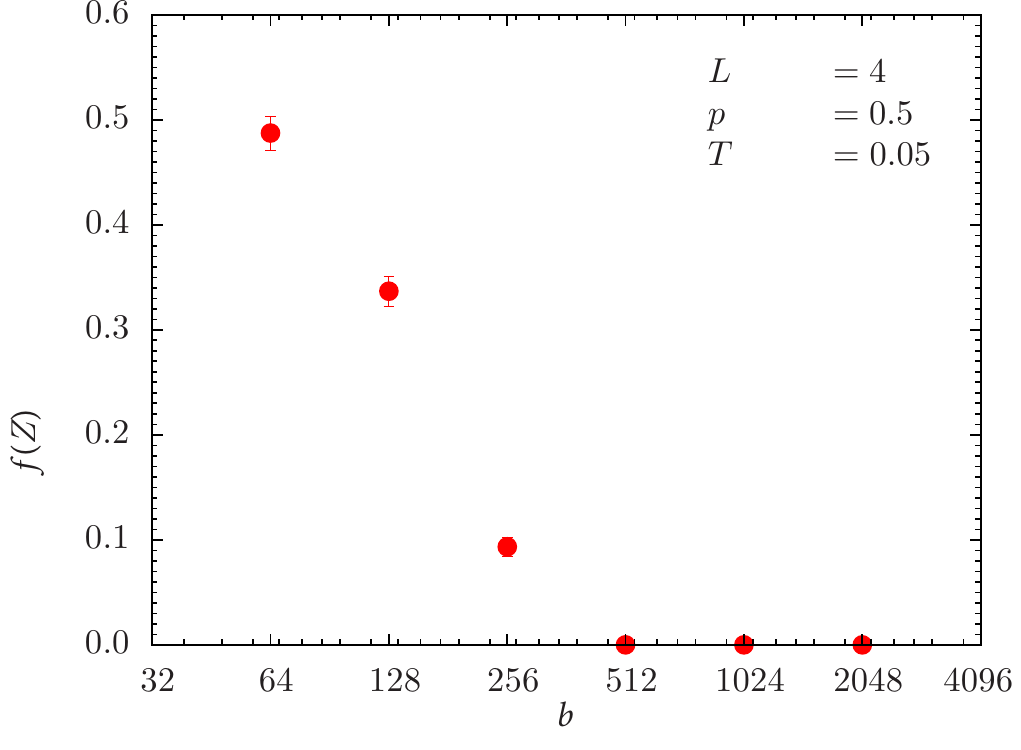}
\vspace*{-1.5em}
\caption{\small
Failure rate of the partition function $Z$ of the two-dimensional 
spin glass as function of precision bits $b$. 
Data averaged over $960$ samples with 
$L=4$, $p=0.5$, $T = 0.05$.
}
\label{fig2}
\end{figure}

To demonstrate that a limited data type precision is the primary
reason for negative partition function values for spin glasses at low
temperature, $960$ samples (linear size $L=4$) with randomly-distributed
bimodal interactions ($p=0.5$) are generated and the partition function
of these samples is calculated with \texttt{double} precision ($b = 64$
bits) and a higher precision ($b=1024$ bits) data type. Figure
\ref{fig1} shows that the failure rate $f(Z)$ of the partition function
$Z$ (i.e., when terms in the sum turn negative) as function of
temperature $T$. It is clear that with \texttt{double} precision the
failure rate increases as temperature decreases. However, the failure
rate vanishes when a precision of $b = 1024$ bits is used.

To further show that a higher precision data type can reduce the failure
rate of partition function for a two-dimensional spin glass, we show the
failure rate as function of date type's precision $b$ in
Fig.~\ref{fig2}.  As the precision increases, the failure rate at fixed
temperature decreases.  These results strongly suggest  that negative
partition function values are  caused by the limited precision of the
data type used in simulations instead of a cut-off parameter $D$
\cite{xie:12}.

\begin{figure}[!tb]
\hspace*{-1.5em}\includegraphics[width=0.85\columnwidth]{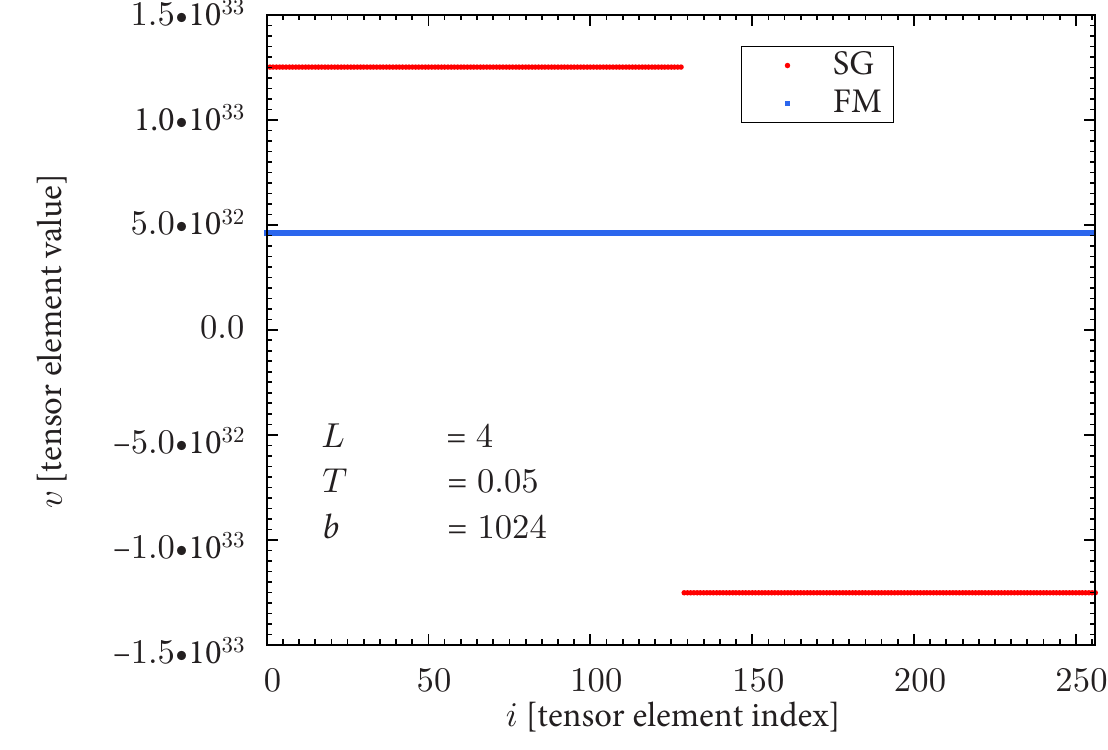}
\vspace*{-1.0em}
\caption{\small
$256$ tensor components of the two-dimensional Edwards-Anderson spin
glass with $p=0.5$, $L=4$ and $T=0.05$ vs the ferromagnetic Ising model
with $L=4$ and $T=0.05$.
}
\label{fig3}
\end{figure}

The success of TRG for the ferromagnetic Ising model implies that there
is an intrinsic difference between the Ising model and a spin glass. To
probe this difference, we plot in Fig.~\ref{fig3} $256$ tensor
components of the contracted tensor $T^{n+1}$ for the two-dimensional
ferromagnetic Ising model, as well as a spin glass. Figure \ref{fig3}
shows that for a spin glass near-cancellation of the positive and
negative tensor components in the partition function requires a higher
precision data type in order to obtain a physical value of the
difference between tensor components. In contrast, for the
ferromagnetic Ising model all components are positive. Therefore
\texttt{double} precision is sufficient to obtain an accurate results.\newline

\noindent {\em Summary} --- By studying the partition function of the
two-dimensional Edwards-Anderson Ising spin-glass model on a square
lattice using the tensor renormalization group method we demonstrate
that the limited precision of the used data type is the culprit for high
failure rates at low temperatures and not a small cut-off parameter as
surmised in Ref.~\cite{xie:12}. To obtain precise partition function
values at low temperature, both a high-precision data type and a large
cut-off parameter are needed. The high precision requirements result in
a sizable numerical overhead when applying tensor renormalization group
methods to spin glasses, because the precision requirements grow for
decreasing temperature or increasing system size. As such, while TRG, in
principle, works for spin glasses, it is an extremely inefficient method
for these models.

{\small \noindent {\em Acknowledgments} --- We would like to thank
Zhiyuan Xie for fruitful discussions. Z.Z.~and H.G.K.~acknowledge
support from the NSF (Grant No.~DMR-1151387).  The research is based
upon work supported by the Office of the Director of National
Intelligence (ODNI), Intelligence Advanced Research Projects Activity
(IARPA), via Interagency Umbrella Agreement IA1-1198. The views and
conclusions contained herein are those of the authors and should not be
interpreted as necessarily representing the official policies or
endorsements, either expressed or implied, of the ODNI, IARPA, or the
U.S.~Government. The U.S.~Government is authorized to reproduce and
distribute reprints for Governmental purposes notwithstanding any
copyright annotation thereon.  We thank the Texas Advanced Computing
Center (TACC) at The University of Texas at Austin and Texas A\&M
University for providing HPC resources.  }

\vspace*{-2.0em}

{\small
\bibliography{refs}

\begin{thebibliography}{8}
\expandafter\ifx\csname natexlab\endcsname\relax\def\natexlab#1{#1}\fi
\expandafter\ifx\csname bibnamefont\endcsname\relax
  \def\bibnamefont#1{#1}\fi
\expandafter\ifx\csname bibfnamefont\endcsname\relax
  \def\bibfnamefont#1{#1}\fi
\expandafter\ifx\csname citenamefont\endcsname\relax
  \def\citenamefont#1{#1}\fi
\expandafter\ifx\csname url\endcsname\relax
  \def\url#1{\texttt{#1}}\fi
\expandafter\ifx\csname urlprefix\endcsname\relax\def\urlprefix{URL }\fi
\providecommand{\bibinfo}[2]{#2}
\providecommand{\eprint}[2][]{\url{#2}}

\bibitem[{\citenamefont{Binder and Young}(1986)}]{binder:86}
\bibinfo{author}{\bibfnamefont{K.}~\bibnamefont{Binder}} \bibnamefont{and}
  \bibinfo{author}{\bibfnamefont{A.~P.} \bibnamefont{Young}},
  \bibinfo{journal}{Rev. Mod. Phys.} \textbf{\bibinfo{volume}{58}},
  \bibinfo{pages}{801} (\bibinfo{year}{1986}).

\bibitem[{\citenamefont{Thomas and Middleton}(2007)}]{thomas:07}
\bibinfo{author}{\bibfnamefont{C.~K.} \bibnamefont{Thomas}} \bibnamefont{and}
  \bibinfo{author}{\bibfnamefont{A.~A.} \bibnamefont{Middleton}},
  \bibinfo{journal}{Phys. Rev. B} \textbf{\bibinfo{volume}{76}},
  \bibinfo{pages}{220406(R)} (\bibinfo{year}{2007}).

\bibitem[{\citenamefont{Wang and Landau}(2001)}]{wang:01}
\bibinfo{author}{\bibfnamefont{F.}~\bibnamefont{Wang}} \bibnamefont{and}
  \bibinfo{author}{\bibfnamefont{D.~P.} \bibnamefont{Landau}},
  \bibinfo{journal}{Phys. Rev. Lett.} \textbf{\bibinfo{volume}{86}},
  \bibinfo{pages}{2050} (\bibinfo{year}{2001}).

\bibitem[{\citenamefont{{Zhu} et~al.}(2015)\citenamefont{{Zhu}, {Ochoa}, and
  {Katzgraber}}}]{zhu:15b}
\bibinfo{author}{\bibfnamefont{Z.}~\bibnamefont{{Zhu}}},
  \bibinfo{author}{\bibfnamefont{A.~J.} \bibnamefont{{Ochoa}}},
  \bibnamefont{and} \bibinfo{author}{\bibfnamefont{H.~G.}
  \bibnamefont{{Katzgraber}}}, \bibinfo{journal}{Phys. Rev. Lett.}
  \textbf{\bibinfo{volume}{115}}, \bibinfo{pages}{077201}
  (\bibinfo{year}{2015}).

\bibitem[{\citenamefont{Hukushima and Nemoto}(1996)}]{hukushima:96}
\bibinfo{author}{\bibfnamefont{K.}~\bibnamefont{Hukushima}} \bibnamefont{and}
  \bibinfo{author}{\bibfnamefont{K.}~\bibnamefont{Nemoto}},
  \bibinfo{journal}{J. Phys. Soc. Jpn.} \textbf{\bibinfo{volume}{65}},
  \bibinfo{pages}{1604} (\bibinfo{year}{1996}).

\bibitem[{\citenamefont{Levin and Nave}(2007)}]{levin:07}
\bibinfo{author}{\bibfnamefont{M.}~\bibnamefont{Levin}} \bibnamefont{and}
  \bibinfo{author}{\bibfnamefont{C.~P.} \bibnamefont{Nave}},
  \bibinfo{journal}{Phys. Rev. Lett.} \textbf{\bibinfo{volume}{99}},
  \bibinfo{pages}{120601} (\bibinfo{year}{2007}).

\bibitem[{\citenamefont{{Xie} et~al.}(2012)\citenamefont{{Xie}, {Chen}, {Qin},
  {Zhu}, {Yang}, and {Xiang}}}]{xie:12}
\bibinfo{author}{\bibfnamefont{Z.~Y.} \bibnamefont{{Xie}}},
  \bibinfo{author}{\bibfnamefont{J.}~\bibnamefont{{Chen}}},
  \bibinfo{author}{\bibfnamefont{M.~P.} \bibnamefont{{Qin}}},
  \bibinfo{author}{\bibfnamefont{J.~W.} \bibnamefont{{Zhu}}},
  \bibinfo{author}{\bibfnamefont{L.~P.} \bibnamefont{{Yang}}},
  \bibnamefont{and} \bibinfo{author}{\bibfnamefont{T.}~\bibnamefont{{Xiang}}},
  \bibinfo{journal}{Phys. Rev. B} \textbf{\bibinfo{volume}{86}},
  \bibinfo{pages}{045139} (\bibinfo{year}{2012}).

\bibitem[{\citenamefont{{Wang} et~al.}(2013)\citenamefont{{Wang}, {Qin}, and
  {Zhou}}}]{wang:13c}
\bibinfo{author}{\bibfnamefont{C.}~\bibnamefont{{Wang}}},
  \bibinfo{author}{\bibfnamefont{S.-M.} \bibnamefont{{Qin}}}, \bibnamefont{and}
  \bibinfo{author}{\bibfnamefont{H.-J.} \bibnamefont{{Zhou}}}
  (\bibinfo{year}{2013}), \bibinfo{note}{(arXiv:cond-mat/1311.6577)}.

\end{thebibliography}
}
\end{document}